\documentclass[%
reprint,
superscriptaddress,
showpacs,preprintnumbers,
showkeys,
 amsmath,amssymb,
 aps,
]{revtex4-2}

\usepackage{graphicx}
\usepackage{dcolumn}
\usepackage{bm}

\def\be{\begin{equation}}
\def\ee{\end{equation}}
\def\bea{\begin{eqnarray}}
\def\eea{\end{eqnarray}}
\newcommand{\ket}[1]{\mbox{$|#1\rangle$}}

\begin{document}

\title{Electrically-Driven and Exponentially-Enhanced  Spin-Photon Interfaces for Quantum Networks}

\author{Fang-Yu Hong}
\affiliation{Zhejiang Key Laboratory of Quantum State Control and  Optical Field Manipulation,Department of Physics,  Zhejiang Sci-Tech University,  Hangzhou, Zhejiang 310018, China}

\date{\today}
\begin{abstract}
    We present an electrically-driven  scheme for  spin-photon quantum interfaces used in quantum networks. Through  modulating the  motion of a nano cantilever with  voltages,  optomechanical coupling and  spin-mechanical coupling  can be exponentially  enhanced simultaneously.    Numerical simulations show that by applying well-designed voltages high-fidelity quantum interface operations such as  generation and  absorption of a single-photon with a known wave packet are within the reach of current techniques.

\end{abstract}


\maketitle
\section{INTRODUCTION}
Nitrogen-vacancy (NV) impurities  in diamond are promising solid-state qubits for emerging quantum computation and quantum communication because of their long spin coherence times  even at room temperature \cite{rhfm,gbal} and fast manipulation speed  \cite{gdfu}, combined  with a level structure allowing for straightforward optical initialization and readout of
the electronic spin state \cite{nmjh} .
Quantum networks comprised of local nodes and quantum channels are
of fundamental  importance  for secure quantum communication \cite{cbgb}, blind computing of both classical and quantum algorithms \cite{abjf},  and modular quantum computing \cite{mcal,ssim}, timing \cite{pkom}, and sensing \cite{zzha}. Quantum interfaces mapping between  optical ``flying" qubits and ``stationary" qubits  is an indispensable component of such a  quantum network. Light-matter quantum interface protocols based on time-depended laser pulses  have been first described and carried out by atomic systems \cite{cirac,adia,wang,hjki,hpsp,kspr,pksh}. 
In each of these experiments or proposals, the quantum interface operations have to be  carried out through precisely controlled laser pulses to  modulate the light-matter interactions which have to be strong enough to reach the strong-coupling regime. Laser pulses are difficult to be produced on-chip and to be manipulated within  a nanoscale region due to its diffraction effect. Thus from the viewpoint
of practical application, it is worth  developing  an electrically driven \cite{seas,kcno,aand, fhru} quantum interfaces for quantum networks 

Recently several proposes \cite{pbli,xfpa,xlhe} have been put forward to exponentially enhance spin-phonon coupling through modulating a spring constant of a mechanical cantilever with a time-dependent voltage. Here we show that by introducing a parametric drive (two-phonon drive)  voltage to modulate a nano cantilever motion we can exponentially enhance   optomechanical (OM) coupling  and  spin-mechanical (SM) coupling simultaneously. With this control at hand we design an electrically-driven scheme to fulfill many kind of  quantum interface functions. Even under the conditions where   the SM coupling and the OM coupling are not satisfied with the strong coupling regime many quantum interface operations such as generating/absorbing a single-photon and establish spin-photon entanglement can still be accomplished with high fidelity.

\begin{figure}[t]
\includegraphics[width=8cm]{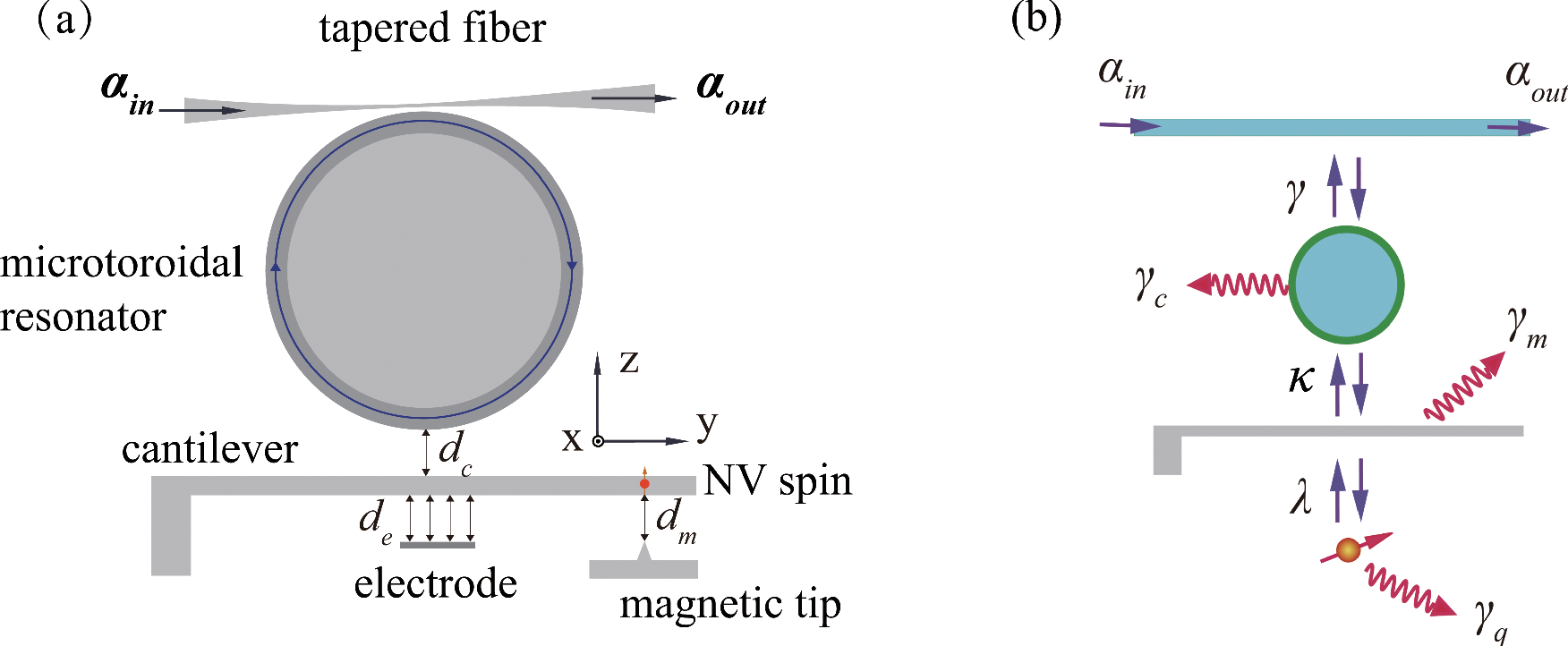}
\caption{\label{fig1}(color online). (a) Schematic setup for a quantum light-matter interface. Driving voltage applied on the electrodes change the nano cantilever's position, which modifies and enhances spin-mechanical interaction and optomechanical interaction simultaneously. Applying voltage pulses can accomplish the generation or absorption of a single-photon wavepacket, resulting in a state transfer or an entanglement distribution between two remote quantum nodes. (b) The interactions among the components of a quantum interface, and corresponding decoherence sources}
\end{figure}

\section{EXPONENTIALLY  ENHANCED  COUPLINGS }
 The prototype quantum interface consists of a high-Q microtoroidal resonator, a nano cantilever with  effective mass $M$  and vibration frequency $ \omega_m$, an optical waveguide (e.g., a tapered fiber), and a NV center embedded in the  cantilever shown in Fig. \ref{fig1}(a). To control the motion of the cantilever, a tunable time-varying voltage $V=V_0+V_p\cos(2\omega_pt)$ is applied on a pair of electrodes of area $A$, one of which is coated on the lower surface of the cantilever \cite{pbli,xlhe}. The spring constant of the cantilever can be regulated through the gradient of the electrostatic force $F=\partial(C_rV^2)/2\partial z$ with the parallel-plate capacitor $C_r=\epsilon A/(d_e+z)$ of the two electrodes, where $z$ denotes the modulated part of the distance relative to the equilibrium $d_e$  between the two electrodes. Under the electrical drive the cantilever motion can be modeled by a Hamiltonian  ($\hbar=1$) \cite{xlhei}
 \be \label{eq10}
 H_\text{m}=\omega_m \hat{a}^\dagger \hat{a} -\Omega_p\cos(2\omega_pt)(\hat{a}+\hat{a}^\dagger)^2,
  \ee
 where 
 \be
 \Omega_p=-\frac{\epsilon AV_0V_p a_0^2}{\hbar d_e^3}
\ee
with the zero-point fluctuation $a_0=\sqrt{\hbar /2M\omega_m}$ and  the corresponding annihilation operator $\hat{a}$.  

 The electronic ground state of the NV spin qubit is  a $S=1$ spin triplet labeled  by  $|m_s\rangle$ with $m_s=0,\pm1$. Two microwave (mw) fields with the same Rabi frequencie $\Omega
$ and the same detuning $\Delta$ drive oscillations between $|0\rangle$ and $|\pm1\rangle$. The NV symmetry axis is assumed to be aligned along the z axis. 
Motion of the cantilever exposes the NV spin to a magnetic field $|B_{\text{tip}}|\backsimeq G_m\hat{z}$ produced by a nearby magnetic tip, which is proportional to the position operator $\hat{z}=a_0(\hat{a}+\hat{a}^\dagger)$ and leads to a Hamiltonian \cite{prab}
\bea \label{eqmq}
H_{\text{mq}}&&=\omega_m\hat{a}^\dagger\hat{a}+\omega_e|e\rangle \langle e|+\omega_{q}|d\rangle \langle d|+(\lambda |g\rangle\langle d| +\lambda_e|d\rangle \langle e|\notag
\\ &&+\text{H.c.})(\hat{a}+\hat{a}^\dagger)-\Omega_p\cos(2\omega_pt)(\hat{a}+\hat{a}^\dagger)^2,
\eea
where $|e\rangle=\cos(\theta)(|-1\rangle+|+1\rangle)/ \sqrt{2}+\sin (\theta)|0\rangle$,  $|g\rangle=\cos(\theta)|0\rangle-\sin (\theta)(|-1\rangle+|+1\rangle)/ \sqrt{2}$,  $|d\rangle=(|-1\rangle-|+1\rangle)/ \sqrt{2}$, $\omega_{q}=\omega_d-\omega_g$, and  $\omega_e=\omega_f-\omega_g$, $\lambda=-\lambda_0\sin (\theta)$, and $\lambda_e=\lambda_0\cos (\theta)$, with $\tan(2\theta)=-\sqrt{2}\Omega/\Delta$, the corresponding eigenfrequencies $\omega_d=-\Delta$,  and  $\omega_{f/g}=(-\Delta \pm \sqrt{\Delta^2+2\Omega^2})/2$. Here $\lambda_0=g_s\mu_BG_ma_0$ with $g_s\simeq2$, and $\mu_B$ is the Bohr magneton. 

Using a rotating transformation $U_0(t)=e^{-iH_0t}$ with $H_0=\omega_p\hat{a}^\dagger \hat{a}$ and dropping the high frequency oscillation and the constant items, the above Hamiltonian \eqref{eqmq} can be rewritten as 
\bea
\tilde{H}_{\text{mq}}&&=\delta_m\hat{a}^\dagger\hat{a}+\omega_e|e\rangle \langle e|+\omega_{q}|d\rangle \langle d|+(\lambda |g\rangle\langle d| +\lambda_e|d\rangle \langle e|\notag
\\ &&+\text{H.c.})(\hat{a}+\hat{a}^\dagger)-\frac{\Omega_p}{2}(\hat{a}^{\dagger 2}+\hat{a}^2),
\eea
where $\delta_m=\omega_m-\omega_p$.

  Meanwhile the cantilever is coupled to a tightly confined optical modes of frequency $\omega_c$ of the microtoroidal cavity through its evanescent field  with the single-photon  OM  coupling rate\cite{gane, kspr, ever}
  \be \label{eq00}
  \kappa_0= a_0(\partial\omega_c/\partial d_c), 
   \ee 
    where $ d_c $ denotes the distance of the cantilever to the toroid.   Since the scattering between two counterpropagating modes can be negligible by positioning the cantilever tangentially to the optical whispering-gallery mode trajectory as in  Fig. \ref{fig1}a \cite{kspr,gane,kspra}, here only a forward circulating cavity mode is considered, which is coupled to the field in the tapered fiber with a constant $\sqrt{\gamma/2\pi}$ \cite{mgap}. The optical field in the tapered fiber can be efficiently coupled to the toroidal microcavity mode which is then coupled back to the forward-propagating field in the fiber with an ideality greater than 99.97\%, which is defined  as the ratio of the amount of power coupled into the desired mode to that coupled into all modes \cite{sstk}. The quantum interface can thus be modeled by a Hamiltonian \cite{pbli,xlhe,smdv, gane, kspr}
    \be
    H=H_{\text{mqc}}+H_{\text{cf}},
    \ee
  where  
\bea \label{eq1}
H_{\text{mqc}}&=&\tilde{H}_{\text{mq}}+\Delta_{c0}\hat{c}^\dagger \hat{c}+\kappa_0\hat{c}^\dagger \hat{c}(\hat{a}+\hat{a}^\dagger)\notag\\
&+&i[\varepsilon(t) \hat{c}^\dagger-\varepsilon(t)^\ast \hat{c}] 
\eea
and 
\be
H_{\text{cf}}=\int_0^\infty \Delta_\omega \hat{f}_\omega^\dagger \hat{f}_\omega d\omega+ \int_0^\infty \left(\sqrt{\frac{\gamma}{2\pi}} (\hat{c}\hat{f}_\omega^\dagger +\hat{c}^\dagger\hat{f}_\omega)\right)d\omega.
\ee
Here $\hat{f}_\omega$ is the annihilation operator for the mode of frequency $\omega$ in the optical channel and $\hat{c}$ is the annihilation operators for the cavity modes of frequency $\omega_c$, and $\varepsilon(t) $ is the slowly varying  strength of an external laser field of frequency $\omega_L$, which coherently excites the cavity field. The detuning  parameters $\Delta_{c0}=\omega_c-\omega_L$ and $ \Delta_\omega=\omega-\omega_L $.  The driving laser field can be sent to the cavity via the tapered fiber shown in Fig.\ref{fig1}. 

In typical experiments the OM coupling $\kappa_0$ is too small to coherently manipulate  the OM system. To achieve a considerable and tunable coupling a strongly driven OM system is widely adopted where the effective OM coupling is enhanced by the coherent field amplitude inside the cavity \cite{kspr, gane, ever, tpal}.  The dissipative dynamics of the cavity field and the cantilever can be treated in terms of quantum Langevin  equations (QLE) \cite{kspra}, 
\be \label{qle1}
\dot{\hat {c}}=-i[\hat {c}, H_{mqc}]-\gamma_t \hat {c}-\sqrt{2\gamma}\hat{f}_{in}(t)-\sqrt{2\gamma_0}\hat{f}_0(t), 
\ee
and
  
\be\label{qle2}
\dot{\hat {b}}=-i[\hat {b}, H_{mqc}]-\gamma_{m0} \hat {b}-\sqrt{\gamma_{m0}}\hat{\xi}(t).
\ee
Here $\gamma_t=\gamma+\gamma_0$ is the total decay rate, $\gamma_0$ is the additional intrinsic losses of the optical cavity mode,  $\hat{f}_{in}(t)$ is the input field operator,  $\hat{f}_0(t)$ is the associated noise operator, $\gamma_{m0}/2$ is the decay rate of the mechanical amplitude, and $\hat{\xi}(t)$ is a thermal white-noise operator satisfying $\langle\hat{\xi}(t)\hat{\xi}^\dagger(t')\rangle=(N_m+1)\delta(t-t')$ and $[\hat{\xi}(t),\hat{\xi}^\dagger(t')]=\delta (t-t')$. Here $N_m$ denotes the Bose occupation number for a mode of frequency $\omega_m$. In the high-temperature case, $k_BT\gg\hbar\omega_m$, the corresponding thermal decoherence rate should be replaced by  $\gamma_m\equiv \gamma_{m0}N_m\approx\frac{k_BT}{\hbar Q}$, where $Q=\omega_m/ \gamma_{m0}$ is the quality factor of the mechanical resonance.

Begining with QLEs \eqref{qle1}  and \eqref{qle2}, we implement a  unitary transformation $\hat{c}\rightarrow \hat{c}+\tilde{c}$ and $\hat{a}\rightarrow \hat{a}+\tilde{a}$  such that the  c number $\tilde{c}$ and $\tilde{a}$ denote the classical mean values of the modes and the new operators $\hat{c}$ and $\hat{a}$ delineate the quantum fluctuations around them. We require all the classical contributions to the transformed QLEs  disappear, which gives 
\be
\dot{\tilde{c}}=-(i\Delta_{c0}+\gamma_t)\tilde{c}-i\kappa_0(\tilde{a}+\tilde{a}^\ast)\tilde{c}+\xi(t),
\ee
 \be
\dot{\tilde{a}}=-(i\delta_m+\frac{\gamma_{m0}}{2})\tilde{a}-i\kappa_0|\tilde{c}|^2+i\Omega_p\tilde{a}^\ast. 
\ee
For $\xi(t)=$const and $\gamma_{m0}\rightarrow 0$ ( $\gamma_m=$const) we have the steady-state solution \
\be
\tilde{a}=\frac{\kappa_0|\tilde{c}|^2}{\Omega_p-\delta_m}
\ee
with $\tilde{c}$ determined from
\be\label{alf}
\tilde{c}=\frac{\xi}{i\Delta_{c0}+\gamma_t+2i\kappa_0^2|\tilde{c}|^2/(\Omega_p-\delta_m)}.
\ee
This formula still approximately holds if $\xi(t)$ is slowly varying compared to the characteristic response time, $\dot{\xi}(t)/ \xi(t)\ll \gamma_t, \Delta_c,\delta_m$ for the case $|\tilde{c}|\kappa_0\ll \Delta_c, \Omega_p-\delta_m$. Through Eq.\eqref{alf} the applied laser power and phase can   be   related to a desired temporal profile $\tilde{c}(t)$. 
 
 In the strong driving regime where $|\tilde{c}(t)|^2\gg 1$ we can linearize  the OM coupling of Eq.\eqref{eq1}: Going back to the displaced operators in the QLEs \eqref{qle1} and  \eqref{qle2} and deleting the classical forces, the terms of order $\kappa_0 $ relative to those of order $\kappa_0 |\tilde{c}| $ and $\kappa_0|\tilde{c}|^2 $ can be omitted. The result of this procedure can also be  obtained by replacing $H_{\text{mqc}}$ with  the linearized OM Hamiltonian 
 
 \be \label{lmqc}
H_{\text{mqc}}^{\text{lin}}=\tilde{H}_{\text{mq}}+\Delta_{c}\hat{c}^\dagger \hat{c}+(\kappa\hat{c}^\dagger +\kappa\hat{c}^\ast )(\hat{a}+\hat{a}^\dagger) 
\ee
 with the laser-enhanced OM coupling $\kappa=\kappa_0\tilde{c}(t)$ and the renormalized cavity detuning $\Delta_c=\Delta_{c0}+2|\kappa(t)|^2/(\Omega_p-\delta_m)$

  To diagonalize the mechanical  mode of Hamiltonian $H^{\text{lin}}=H_{\text{mqc}}^{\text{lin}}+H_\text{cf}$ we take the unitary Bogoliubov transformation $U_S(r(t))=\exp[r(t)(\hat{a}^2-\hat{a}^{\dagger 2})/2]$ with the squeezing parameter $r(t)$ defined by $\tanh 2r(t)=\Omega_p(t)/\delta_m(t)$, leading to the  Hamiltonian 
\bea \label{eq4}
H^{\text{S}}&=&\omega_{e}|e\rangle \langle e| +\omega_{q}|d\rangle \langle d|+\Delta_m\hat{a}^\dagger \hat{a}+\Delta_c\hat{c}^\dagger \hat{c}\notag
\\&+&\int_0^\infty\Delta_\omega \hat{f}_\omega^\dagger \hat{f}_\omega d\omega +\int_0^\infty \left(\sqrt{\frac{\gamma}{2\pi}} (\hat{c}\hat{f}_\omega^\dagger +\hat{c}^\dagger\hat{f}_\omega)\right)d\omega\notag
\\&+&(\lambda_{\text{ef}}|g\rangle \langle d| +\lambda_{e}e^{r(t)} |d\rangle \langle e|+\text{H.c.})(\hat{a}+\hat{a}^\dagger)\notag\\
&+&(\kappa_{\text{ef}}\hat{c}^\dagger+\kappa_{\text{ef}} \hat{c})(\hat{a}+\hat{a}^\dagger)+\frac{i}{2}\dot{r}(t)(\hat{a}^2-\hat{a}^{\dagger 2}),
\eea
where $\Delta_m=\delta_m/\cosh 2r(t)$, $\lambda_{\text{ef}}=\lambda e^{r(t)}$, and $\kappa_{\text{ef}}=\kappa e^{r(t)}$ with $\lambda$ and $\kappa$ assuming to be real.
Applying a unitary transformation $U=\exp(i\pi |d\rangle\langle d|/2+i\pi\hat{c}^\dagger\hat{c}/2)$ and the rotating-wave approximation (RWA) with the resonance condition $\Delta'_m\equiv-\Delta_m\thickapprox \Delta_c\thickapprox\omega_q\neq\omega_e$ Hamiltonian \eqref{eq4} reduces to 
\bea \label{eq5}
&H&^{\text{S}}_{\text{rwa}}=\frac{\omega_q}{2}\hat{\sigma}_z+\Delta_m\hat{a}^\dagger \hat{a}+\Delta_c\hat{c}^\dagger \hat{c}+\int_0^\infty \Delta_\omega \hat{f}_\omega^\dagger \hat{f}_\omega d\omega+\notag
\\&&i\bigg( \lambda_{\text{ef}}\hat{\sigma}_+\hat{a}^\dagger  +\kappa_{\text{ef}}\hat{c}^\dagger\hat{a}^\dagger+\sqrt{\frac{\gamma}{2\pi}} \int_0^\infty   \hat{c}^\dagger\hat{f}_\omega d\omega-\text{H.c.}\bigg),
\eea
where $\hat{\sigma}_z$ is the Pauli operator and $\hat{\sigma}_+=|d\rangle\langle g|$ is the raising operator for the spin qubit. Note that we work in the regime where $\Delta_m<0$, for the sake of convenience  Hamiltonian \eqref{eq5} can be rewritten as 
\bea \label{eq5a}
&&H^{\text{S}}_{\text{f}}=\frac{\omega_q}{2}\hat{\sigma}_z+\Delta'_m\hat{a}_m^\dagger \hat{a}_m+\Delta_c\hat{c}^\dagger \hat{c}+\int_0^\infty \Delta_\omega \hat{f}_\omega^\dagger \hat{f}_\omega d\omega+\notag
\\&&i\left( \lambda_{\text{ef}}\hat{\sigma}_+\hat{a}_m + \kappa_{\text{ef}}\hat{c}^\dagger\hat{a}_m+ \sqrt{\frac{\gamma}{2\pi}} \int_0^\infty  \hat{c}^\dagger\hat{f}_\omega d\omega -\text{H.c.}\right),
\eea
  where $\hat{a}_m=\hat{a}^\dagger$ and the constant energy has been omitted. The coupling strength $\lambda_{\text{ef}}$ and $\kappa_{\text{ef}}$ are both exponentially enhanced simultaneously because of the modulating parametric driving  voltage applied on the cantilever. 
  
\begin{figure}[t]
\includegraphics[width=8cm]{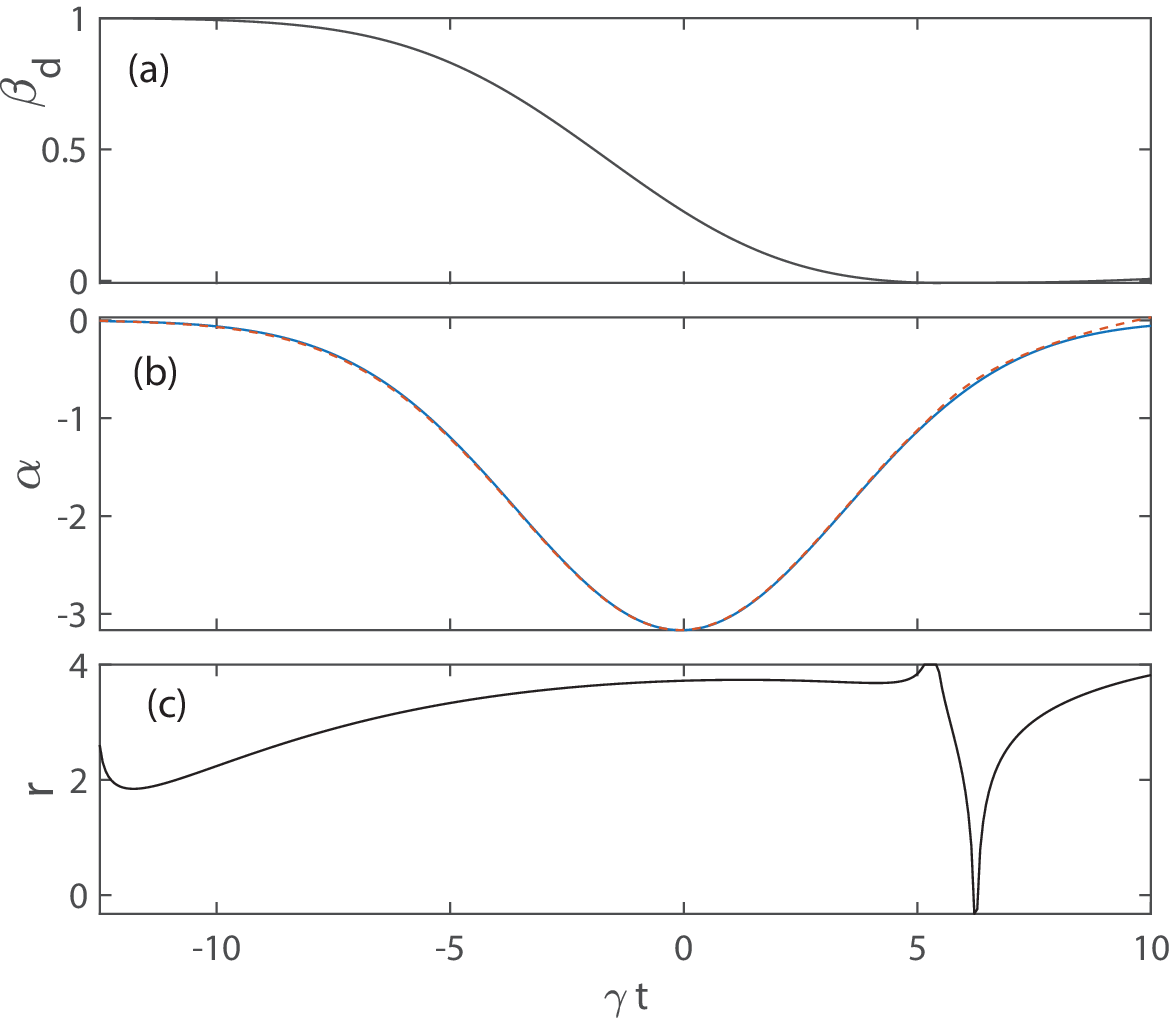}
\caption{\label{fig2} (Color online). Numerical simulation of generating a single-photon. The parameters used are $\gamma/2\pi=10$MHz, $\kappa/2\pi=0.09$MHz, $\lambda/2\pi=0.05$MHz, $\gamma_c=0.01\gamma$, $\gamma_m=0.01\lambda$,  and $\gamma_q/2\pi=2$kHz. (a) The evolution of state amplitude $\beta_{d}(t)$. (b) The ideal single-photon wavepacket $\alpha_{\text{out}}^{\text{ideal}}(t)=\tilde{\alpha}(t)$ (solid line) and the generated single-photon wavepacket $\alpha_{\text{out}}$ (dashed line). (c) The driving squeezing parameter $r$.}
\end{figure}

\section{QUANTUM INTERFACE CONTROL SCHEME}
Under the RWA which holds if $ \lambda_{\text{ef}},\kappa_{\text{ef}}, \dot{r}/2\ll \omega_q+\Delta_c$ the probability of generating more than one exciton is negligible. Throughout the whole process of interconverting  local and flying qubits  in the quantum interface, the state $\ket{g,0,0}\ket{vac}$ does not evolve into the subspace spanned by basis $ \ket{d,0,0}\ket{vac} $,  $ \ket{g,1,0}\ket{vac} $, $ \ket{g,0,1}\ket{vac} $, and $ \hat{f}_\omega^\dagger\ket{g,0,0}\ket{vac}$, where $\ket{vac}$ is the vacuum state of the flying qubit, and in notations $\ket{u,j,k}$, $u=g, d$ denotes the stationary qubit states, $j$, $k$ denote the number of excitations in the mechanical and cavity mode, respectively.  Hence, in the interaction picture the evolution of the system are in the form  $ \ket{\Psi}=C_g\ket{g,0,0}\ket{vac}+C_d\ket{\Psi^d(t)}$, where
\bea
\ket{\Psi^d(t)}&=&\int_0^\infty d\omega\alpha_\omega \hat{f}_\omega^\dagger\ket{g,0,0}\ket{vac}+\beta_d\ket{d,0,0}\ket{vac} \nonumber\\
&+&\beta_m\ket{g,1,0}\ket{vac}+\beta_c\ket{g,0,1}\ket{vac}.
\eea

Under the Hamiltonian given in Eq.\eqref{eq5a},  the Schr\"{o}dinger equations for the state amplitudes of the quantum interface system  in the interaction picture can be derived as
\begin{subequations}\label{e6}
\bea
\dot{\beta}_d&=&\lambda_{\text{ef}}\beta_m\text{e}^{-i(\Delta'_m-\omega_q)t},\label{q6a}\\
\dot{\beta}_m&=&-\lambda_{\text{ef}}\beta_d\text{e}^{i(\Delta'_m-\omega_q)t}
-\kappa_{\text{ef}}\beta_c\text{e}^{i(\Delta'_m-\Delta_c)t}, \label{q6b}\\
\dot{\beta}_c&=&\kappa_{\text{ef}}\beta_m\text{e}^{-i(\Delta'_m-\Delta_c)t}\notag
\\&&+\sqrt{\frac{\gamma}{2\pi}}\int_{0}^{\infty} \alpha_\omega \text{e}^{-i(\Delta_\omega-\Delta_c)t}d\omega  \label{q6c},\\
\dot{\alpha}_\omega&=&-\sqrt{\frac{\gamma}{2\pi}}\beta_c\text{e}^{i(\Delta_\omega-\Delta_c)t}.\label{q6d}
\eea
\end{subequations}
Integrating equation \eqref{q6d} yields 
\be\label{e1}
\alpha_\omega (t)=\alpha_\omega (t_0)-\sqrt{\frac{\gamma}{2\pi}}\int_{t_0}^{t}\text{e}^{i(\Delta_\omega-\Delta_c)t'}\beta_c(t')dt',
\ee
or
\be\label{e2}
\alpha_\omega (t)=\alpha_\omega (t_1)+\sqrt{\frac{\gamma}{2\pi}}\int_{t}^{t_1}\text{e}^{i(\Delta_\omega-\Delta_c)t'}\beta_c(t')dt',
\ee
where $t_0\rightarrow-\infty$ and $t_1\rightarrow\infty$ denote the remote past and remote future respectively when the incoming/outgoing photon wavepackets are not influenced by the quantum interface. 

By substituting equation \eqref{e1} or \eqref{e2} into equation \eqref{q6c} and applying the Weisskopf-Wigner approximation \cite{mosc}  the Schr\"{o}dinger equations \eqref{e6} can be rewritten as
\begin{subequations}\label{eq6}
\bea
\dot{\beta}_d&=&\lambda_{\text{ef}}\beta_m\text{e}^{-i(\Delta'_m-\omega_q)t},\label{eq6a}\\
\dot{\beta}_m&=&-\lambda_{\text{ef}}\beta_d\text{e}^{i(\Delta'_m-\omega_q)t}
-\kappa_{\text{ef}}\beta_c\text{e}^{i(\Delta'_m-\Delta_c)t}, \label{eq6b}\\
\dot{\beta}_c&=&\kappa_{\text{ef}}\beta_m\text{e}^{-i(\Delta'_m-\Delta_c)t}-\sqrt{\gamma} \alpha_{in}(t)-\frac{\gamma}{2}\beta_c \label{eq6c},\\
&=&\kappa_{\text{ef}}\beta_m\text{e}^{-i(\Delta'_m-\Delta_c)t}-\sqrt{\gamma} \alpha_{out}(t)+\frac{\gamma}{2}\beta_c,\label{eq6d}
\eea
\end{subequations}
where $\alpha_{in}(t)\equiv\int d\omega\alpha_\omega(t_0)\text{e}^{i(\Delta_c-\Delta_\omega)t}/\sqrt{2\pi}$ with $t_0\rightarrow-\infty$ and $\alpha_{out}(t)\equiv\int d\omega\alpha_\omega(t_1)\text{e}^{i(\Delta_c-\Delta_\omega)t}/\sqrt{2\pi}$ with $t_1\rightarrow+\infty$ are the incoming and outgoing photon pulses in the quantum channel, respectively.

According to equations \eqref{eq6c} and \eqref{eq6d}  the solution for $\beta_c$ is
\be\label{eq7}
\beta_c(t)=\frac{1}{\sqrt{\gamma}}\bigg(\alpha_{out}(t)-\alpha_{in}(t)\bigg).
\ee

Substituting equation \eqref{eq6c} into equation \eqref{eq6a} gives 
\be\label{eq8}
\dot{\beta}_d(t)=\frac{\lambda}{\kappa}\bigg(\dot{\beta}_c+\sqrt{\gamma}\alpha_{in}(t)+\frac{\gamma}{2}\beta_c\bigg)\text{e}^{i(\omega_q-\Delta_c)t}.
\ee

From equations \eqref{eq6} the normalization condition can be derived as follows
\be\label{eq10}
\frac{d}{dt}\left(|\beta_d|^2+|\beta_m|^2+|\beta_c|^2\right)=|\alpha_{in}(t)|^2-
|\alpha_{out}(t)|^2.
\ee
From Eq.\eqref{eq6b} we have 
\bea\label{eq}
|\beta_m|^2\frac{d}{dt} \theta_m&=&i\bigg(-|\dot{ \beta}_m | | \beta_m |+\lambda_{ef}\beta_d\beta_m^\ast\text{e}^{i(\Delta'_m-\omega_q)t} \notag
\\ &+& \kappa_{ef}\beta_c\beta_m^\ast\text{e}^{i(\Delta'_m-\Delta_c)t} \bigg) 
\eea
with $\theta_m=\text{arg}(\beta_m)$. 
From Eq. \eqref{eq} and Eq. \eqref{eq6a} we have 
\be \label{eq10a}
\frac{d}{dt}  \theta_m=\frac{i}{2|\beta_m|^2}\left(\beta_d\dot{\beta_d}^\ast+\frac{\kappa}{\lambda}\beta_c\dot{\beta_d}^\ast \text{e}^{i(\Delta_c-\omega_q)t} -\text{c.c.}\right),
\ee
where `c.c.' denotes complex conjugate.

At last from equation \eqref{eq6c}, we get the squeezing parameter $r$ 
\be\label{eq9}
\text{e}^r=\frac{1}{\kappa\beta_m}\bigg(\dot{\beta}_c+\sqrt{\gamma}\alpha_{in}(t)+\frac{\gamma}{2}\beta_c\bigg)\text{e}^{i(\Delta'_m-\Delta_c)t}.
\ee

\begin{figure}[t]
\includegraphics[width=8cm]{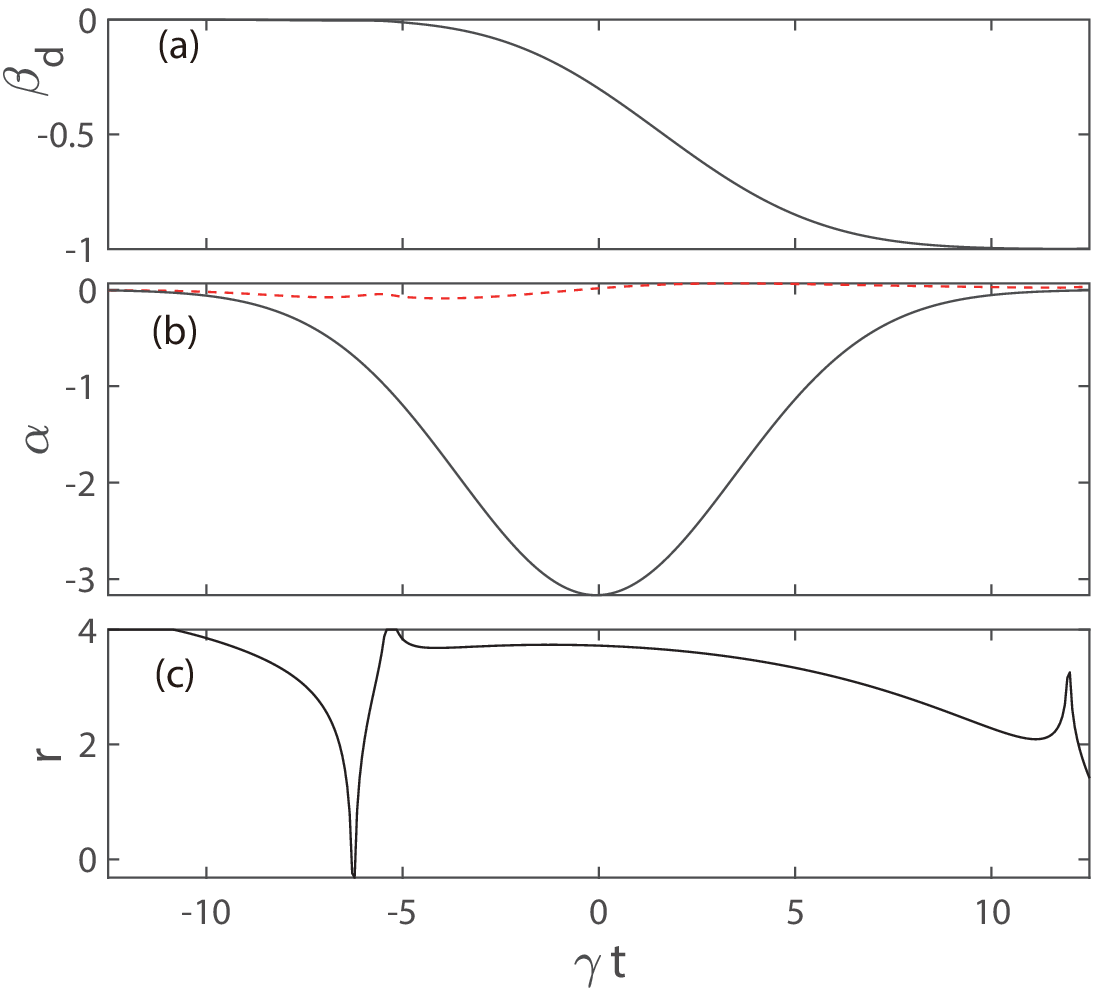}
\caption{\label{fig3} (Color online). Numerical simulation of absorbing a single-photon wavepacket. The parameters used are as those in Fig. \ref{fig2}. (a) The evolution of state amplitude $\beta_{d}(t)$. (b) The incoming single-photon wavepacket $\alpha_{\text{in}}(t)=\tilde{\alpha}(t)$ (solid line) and the reflected photon wavepacked (dashed line). (c) The driving squeezing parameter $r$.}
\end{figure}

 For simplicity $\alpha_{out}(t)$ and $\alpha_{in}(t)$ are both assumed to be real and the resonance condition $\Delta'_m=\Delta_c=\omega_q $ is satisfied. The quantum interface can be designed in this way:  first, the wave packets of the outgoing/incoming  single-photon
are arbitrarily assigned only if they are sufficiently smooth; the evolvement of cavity mode $\beta_c(t)$ is then determined according to Eq. \eqref{eq7}; third, the state amplitudes $\beta_d$ can be solved from Eq. \eqref{eq8}; next the evolution of $\beta_{m}$ can be obtained from equations \eqref{eq10} and \eqref{eq10a}; and the squeezing parameter  $r$ can be determined according to Eq. \eqref{eq9} with $\Delta'_m=\Delta_c$; at last we choose appropriate parameters such as $\omega_p$, $\Delta$, and $\Omega$   to satisfy the resonance condition 
\be
\frac{\omega_p(t)-\omega_m}{\cosh (2r(t))}=\Delta_c=\omega_q.
\ee

For the sending node of a quantum network, the initial conditions are $\alpha_{in}(t)=0$, $\beta_d(t_0)=1$, $\beta_m(t_0)=0$, $\beta_c(t_0)=0$. The outgoing single-photon wave packet can contain average $\sin^2\varphi$ photon with a single-photon wavepacket $\tilde{\alpha}(t)$: $\int_{t_0}^{t_1} dt |\alpha_{out}(t)|^2=\sin^2\varphi\int_{t_0}^{t_1} dt |\tilde{\alpha}(t)|^2=\sin^2\varphi$. At the remote future time $t_1\rightarrow+\infty$, the photon generation process is completed, we have $\beta_m(t_1)=0$, $\beta_c(t_1)=0$, and $\beta_d(t_1)=\cos\varphi\text{e}^{\phi}$ with the phase $ \phi$ determined by Eq. \eqref{eq8}, The most general form of the photon generation process in the quantum interface can be described by \cite{wang}
\bea\label{eq11}
&C_g&\ket{g,0,0}\ket{vac}+C_d\ket{d,0,0}\ket{vac}\xrightarrow {r(t)}C_g\ket{g,0,0}\ket{vac}\notag\\
&+&C_d[\text{e}^{\phi}\cos\varphi\ket{d,0,0}\ket{vac}+\sin\varphi\ket{g,0,0}
\ket{\tilde{\alpha}(t)}].
\eea
If $\varphi=\pi/2$, Eq. \eqref{eq11} is reduced to  
\bea\label{eq12}
&C_g&\ket{g,0,0}\ket{vac}+C_d\ket{d,0,0}\ket{vac}\xrightarrow{r(t)}\ket{g,0,0}[C_g\ket{vac}\notag\\
&+&C_d\ket{\tilde{\alpha}(t)}],
\eea
mapping the stationary qubit state onto the flying qubit. Further, if initially the qubit is in  state $\ket{d}$, then this mapping operation can work as the deterministic generation of a single-photon with any desired pulse shape $\tilde{\alpha}(t)$. If $\varphi<\pi/2$, this sending node can also work as generating entanglement between the stationary qubit and the flying qubit:
\bea\label{eq13}\label{eq14}
\ket{d,0,0}\ket{vac}\xrightarrow{r(t)}\text{e}^{\phi}\cos\varphi\ket{d,0,0}\ket{vac}\notag\\
+\sin\varphi\ket{g,0,0}\ket{\tilde{\alpha}(t)}.
\eea

The receiving process is basically the time reversal of the
sending process under the condition $\varphi=\pi/2$ \cite{wang}. With the qubit initially in state
$\ket{g}$ and the incoming flying qubit in state $C_g\ket{vac}\notag+C_d\ket{\tilde{\alpha}(t)}$, the mapping transformation is
\bea\label{eq15}
\ket{g,0,0}(C_g\ket{vac}\notag+C_d\ket{\tilde{\alpha}(t)})\xrightarrow{r(t)}\notag\\
(C_g\ket{g,0,0}\ket{vac}-C_d\ket{d,0,0})\ket{vac},
\eea
followed by a unitary transformation $U_2=\begin{bmatrix}
1&0\\0&-1
\end{bmatrix}$ on the spin qubit  in the receiving node to accomplish a state transfer $C_g\ket{g}_1+C_d\ket{d}_1\rightarrow C_g\ket{g}_2+C_d\ket{d}_2$ with subscript $i=1,2$ represents the spin qubit in the sending/receiving node, respectively.
By combining the sending and receiving process, the transfer
of an arbitrary unknown qubit state  from one node to another can be accomplished.
When two neighboring nodes carry out state transfer operations followed by  receiving state operation at the same time, the two qubits states are swapped. If $\varphi<\pi/2$ for the sending node, the joint operation of the sending and receiving process can generate an entanglement between two remote nodes by transformation
\be\label{eq16}
\ket{d}_1\ket{vac}\ket{g}_2\xrightarrow[r_2(t)]{r_1(t)}\left(\text{e}^{\phi}\cos\varphi\ket{d}_1\ket{g}_2
-\sin\varphi\ket{g}_1\ket{d}_2\right)\ket{vac},
\ee
with the corresponding  squeezing parameters $r_i(t) (i=1,2)$  for the sending / receiving node, respectively. 

\begin{figure}[t]
\includegraphics[width=8cm]{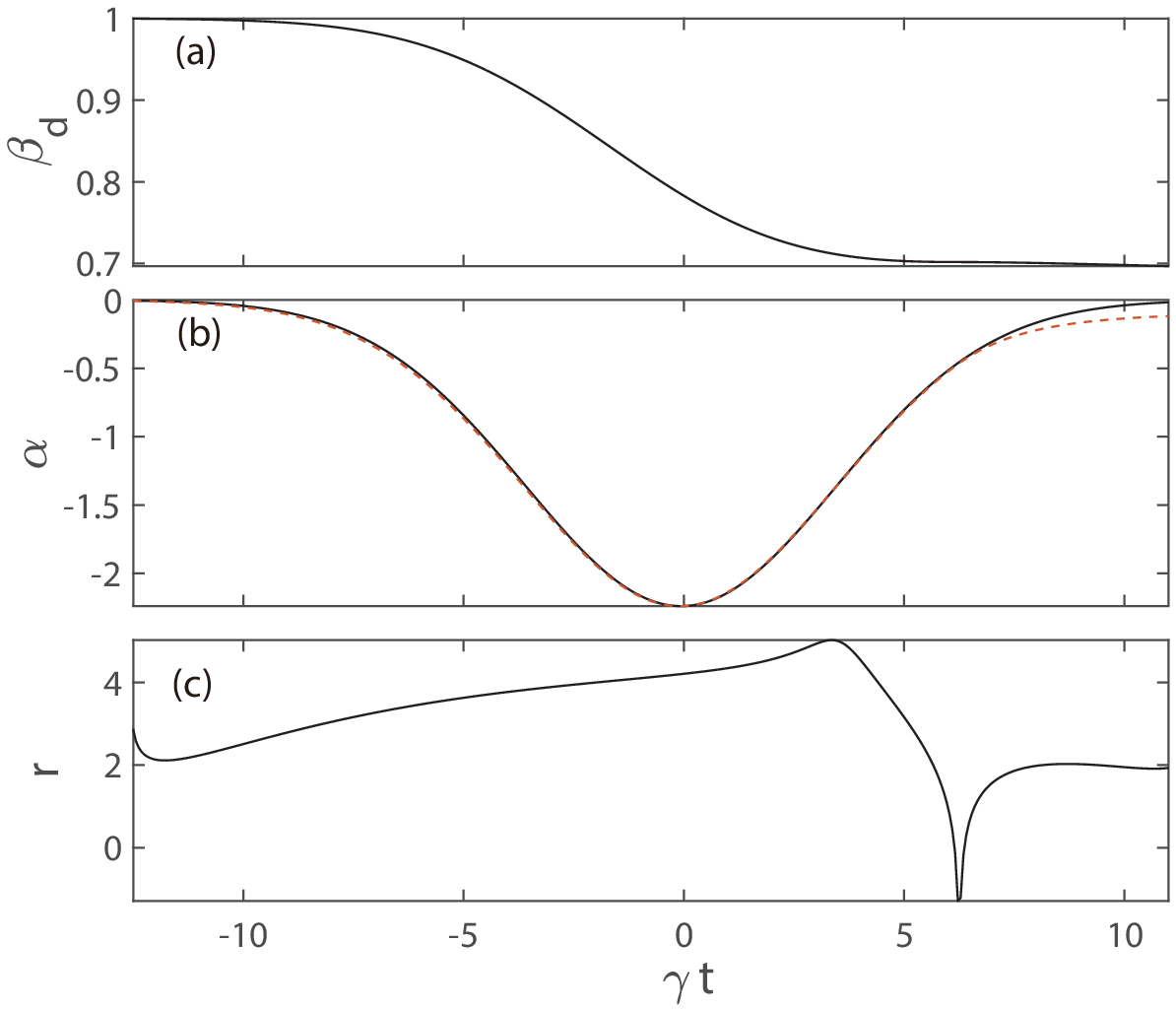}
\caption{\label{fig4} (Color online). Establishing entanglement $|\psi\rangle=\frac{1}{\sqrt{2}}(|d\rangle|vac\rangle +|g\rangle|\tilde{\alpha}(t)\rangle) $ between a spin qubit and a single-photon. The corresponding parameters remain the same as those in figure \ref{fig2} except $\lambda/2\pi=0.02$ MHz. (a) The evolution of state amplitude $\beta_{d}(t)$. (b) The ideal single-photon wavepacket $\alpha_{\text{out}}^{\text{ideal}}(t)=\frac{1}{\sqrt{2}}\tilde{\alpha}(t)$ (solid line) and the generated single-photon wavepacket $\alpha_{\text{out}}$ (dashed line). (c) The driving squeezing parameter $r$.}
\end{figure}

\section{NUMERICAL SIMULATION AND DISCUSSION}
Now we discuss the effects arising from some inevitable decoherence sources. In the high-temperature situation where $k_B T \gg \hbar\omega_m$, the effective mechanical dissipation rate  $\gamma_m\approx \frac{k_BT}{\hbar Q}$ \cite{kspr}. When the coupling rates $\lambda_{\text{ef}}$ and $\kappa_{\text{ef}}$ are exponentially enhanced, so does the effective cantilever decoherence rate. But the harmful effect of amplified mechanical noises can be circumvent  by applying the experimentally accomplished  dissipative squeezing technique \cite{mlnd,ewcl,jped}, in which the Bogoliubov mode can  be kept in its ground state \cite{stsm}.

Taking into account the decoherence including the photon leakage into free space (with rate $\gamma_c$) and that from the cantilever and the spin qubit (with rate $\gamma_q$)  the Schr\"{o}dinger equations \eqref{eq6} should be replaced by
\begin{subequations}\label{eq17}
\bea
\dot{\beta}_d&=&\lambda_{\text{ef}}\beta_m-\frac{\gamma_q}{2}\beta_d,\label{eq17a}\\
\dot{\beta}_m&=&-\lambda_{\text{ef}}\beta_d
-\kappa_{\text{ef}}\beta_c-\frac{\gamma_m}{2}\beta_m, \label{eq17b}\\
\dot{\beta}_c&=&\kappa_{\text{ef}}\beta_m-\sqrt{\gamma} \alpha_{in}(t)-(\frac{\gamma}{2}+\frac{\gamma_c}{2})\beta_c, \label{eq17c}\\
&=&\kappa_{\text{ef}}\beta_m-\sqrt{\gamma} \alpha_{out}(t)+(\frac{\gamma}{2}-\frac{\gamma_c}{2})\beta_c,\label{eq17d}
\eea
\end{subequations}
where for the sake of simplicity, we have assumed the resonance conditions $\Delta'_m=\omega_q=\Delta_c$.

The scheme is to design the squeezing parameter $r(t)$ according to equation \eqref{eq9}  through assigning  target single-photon wavepacket $\alpha_{in}(t)$ or $\alpha_{out}(t)$, and use  it to drive equations \eqref{eq17} under the real situation. First we simulate generating a single-photon wavepacket $\alpha_{\text{out}}^{\text{ideal}}(t)=\tilde{\alpha}(t)\equiv-\text{exp}(-(\gamma t/5)^2)$ with normalization understood, under the driving voltage applied on the  electrodes, which is dependent on the squeezing parameter $r$ shown in Fig. \ref{fig2}(c),  the evolution of the whole quantum interface can be obtained, here we plot the evolution of amplitude of the state $\beta_d$ in Fig. \ref{fig2}(a) and the generated single-photon wavepacket $\alpha_{\text{out}}$ in Fig. \ref{fig2}(b) (dashed line). In the simulation we have used  the corresponding  parameters $\lambda/2\pi=0.05$ MHz, $\kappa/2\pi=0.09$MHz, $\gamma/2\pi=10$MHz, $\gamma_c=0.01\gamma$, $\gamma_m=0.01\lambda$,  and $\lambda_q/2\pi=2$kHz. The single photon generation fidelity $\langle\alpha_{\text{out}}^{\text{ideal}}|\alpha_{\text{out}}\rangle=0.9879$. If we increase the mechanical decoherence rate to $\gamma_m=\lambda$, the fidelity  slightly decrease to 0.9760.  

Next we simulate absorbing a single-photon wavepacket $\alpha_{\text{in}}^{\text{ideal}}(t)=\tilde{\alpha}(t)$ shown in Fig. \ref{fig3}. The photon absorption fidelity is $ \beta_d^{\text{ideal}}(t_1)\beta_d(t_1)= 0.9869$ with $\beta_d^{\text{ideal}}(t_1)=-1$ and the probability of an incoming single-photon  wavepacket being reflected  is $\int |\alpha_{\text{out}}(t)| ^2dt=0.11\% $. Figure \ref{fig4} shows establishing entanglement between a spin qubit and a flying qubit through generating a photon wavepacket $\alpha_{\text{out}}^{\text{ideal}}(t)=\frac{1}{\sqrt{2}}\tilde{\alpha}(t)$.  The fidelity of establishing spin-photon entanglement is $\langle\alpha_{\text{out}}^{\text{ideal}}|\alpha_{\text{out}}\rangle
+\beta_d^{\text{ideal}}(t_1)\beta_d(t_1)=0.9841$ with $\beta_d^{\text{ideal}}(t_1)=\frac{1}{\sqrt{2}}$. All those operations can be finished within $0.4\mu$s. Note that in all these simulations, $\gamma_c=0.01\gamma>\lambda,\kappa$, thus we are working in the weak coupling regime, yet we may still accomplish high-fidelity quantum interface operations.    

In the above simulations we have assumed exact knowledge of the parameters $r(t)$, $\lambda$, $\kappa$, and $\gamma$. But in fact, there may be various errors in those parameters because of all kind of imperfections. Thus we need to evaluate the influence of these errors on the fidelity of quantum interface operations. Table \ref{tab1} shows the effects of the unknown errors in the various parameters on the fidelity of generating a single-photon wavepacket $\tilde{\alpha}(t)$, absorbing wavepacket $\tilde{\alpha}(t)$, and establishing entanglement $|\psi\rangle=\frac{1}{\sqrt{2}}(|d\rangle|vac\rangle +|g\rangle|\tilde{\alpha}(t)\rangle) $.

 To examine the experimental feasibility of this proposal, considering the cantilever with the dimensions ($l=100,w=0.02,t=0.01$) $\mu$m used in the literature \cite{xlhei} we  adopt a  silicon cantilever with a less demanding dimensions ($l=50,w=0.02,t=0.01$) $\mu$m,  which has a fundamental frequency $\omega_m\sim3.516 \times (t/l^2)\sqrt{E/12\rho}\sim 2\pi \times 4.83 $ kHz and $a_0=\sqrt{\hbar/2M\omega_m}\sim 1.73\times 10^{-11}$m, with the mass density $\rho\sim2.33 \times 10^3 $kg/m$^3$, Young's modulus $E\sim 1.3\times 10^{11}$ Pa, and the effective mass $M\sim \rho lwt/4$ \cite{pbli, xlhei}. A magnetic gradient of $G_m\approx 1.4 \times 10^6$ T/m  generated by a magnetic tip with size of $\sim 100$ nm and homogenous magnetization $M_m\approx2.3 \times 10^6$ T/$\mu_0$  at the distance $d_m\approx 25$ nm away from the tip was reported in \cite{hjma}. Then we have  a coupling strength $\lambda_0/2\pi=g_s\mu_BG_ma_0/2\pi \approx0.68$ MHz. We set $\theta\approx0.22 \pi$ at the ``sweet spot" and obtain $\lambda/2\pi \approx-0.43$ MHz \cite{prab, ffun}. Note that in the numerical simulations we assume  $\lambda/2\pi=0.05$ MHz in Fig.\ref{fig2}, \ref{fig3} and $\lambda/2\pi=0.02$ MHz in Fig.\ref{fig4} . Here this minus ``$-$" is not important,  we can apply a unitary transformation $U=\exp(i\pi |d\rangle\langle d|)$ on the Hamiltonian (Eq. \ref{eq5a}) to absorb it into $\hat{\sigma}_+$.
  
  For an environmental temperature $10$ mK and a quality factor $Q\sim10^6$ the effective mechanical disspation rate 
 $\frac{k_BT}{\hbar Q}\approx2\pi \times 0.21$ KHz. For the NV spin qubit we assume $\omega_q/2\pi=0.7$ GHz, $\Delta/2\pi=-0.11$GHz, and $\Omega/2\pi=0.41$ GHz  to meet the ``sweet spot" condition $\theta\approx0.22 \pi$ where the quadratic corrections to the transition frequency $\omega_q$ vanish and results in a spin decoherence rate $\gamma_q/2\pi\approx0.6\times\delta_n^4/(2\pi \omega_q^3) < 1$Hz  with $\delta_n\simeq1$ MHz \cite{prab}.  The drive amplitude can reach  $-\Omega_p/ 2\pi  \approx 8000$ GHz with the parameters $V_0=10$V, $V_p=2$V, $A=5\mu$m$\times0.02\mu$m, and $d_e=0.01\mu$m; while the maximum drive amplitude occurred in the simulation  $-\Omega_p/ 2\pi  \approx 7709$ GHz for the maximum squeezing parameter $r=5$ (Fig. \ref{fig4}) and $\omega_q/2\pi=0.7$GHz. Concerning the optical parameters $\gamma_c/2\pi=0.01\gamma$ and  $\gamma/2\pi=10 $MHz correspond to optical quality factor $Q_c\equiv\pi c/(\gamma\lambda_p )=1.76\times10^7$ with $\lambda_p=852$ nm \cite{kspr, sstk, sstkk}. Compared with $Q_c\geq2\times10^9$ adopted in the seminal literature \cite{kspr} to obtain  state transfer fidelities $F\approx0.85$, this scheme significantly relaxes the requirement on the cavity quality.  As for the optomechanical coupling rate $\kappa/2\pi=1.8$ MHz has been reported in the literature \cite{ever}.
 
 Now we discuss the RWA conditions. For the parameters $\dot{r}/2/2\pi\sim5$ MHz (Fig.\ref{fig2},\ref{fig3},\ref{fig4}),  $\kappa_{\text{ef}}/2\pi=13.4$ MHz, $\lambda_{\text{ef}}/2\pi=2.97$ MHz for $r=5$, and $\omega_q/2\pi=\Delta_c/2\pi=0.7$ GHz, the RWA conditions  $ \lambda_{\text{ef}},\kappa_{\text{ef}}, \dot{r}/2\ll \omega_q+\Delta_c$ are satisfied very well.
 
\begin{table}
\caption{\label{tab1}Effect of errors in parameters on the fidelities  
of generating a single-photon wavepacket $\tilde{\alpha}(t)$, absorbing such a wavepacket, and establishing entanglement $|\psi\rangle$  . The evaluations are carried out by adding $+(-)10\% $ to the squeezing parameter $r(t)$, $\lambda$, $\kappa$, or $\gamma$ respectively in the numerical simulations with other parameters remained the same as in figures \ref{fig2}, \ref{fig3}, \ref{fig4}. }
\begin{ruledtabular}
\begin{tabular}{cccccc}
& &10\% $r(t)$ &10\% $\lambda$&10\% $\kappa$ &10\%$\gamma$ \\
&no error&error& error& error& error\\
\hline
Generate &0.9862&0.8424&0.9801&0.9812&0.9873\\
Absorb&0.9869&0.8280&0.8975&0.9062&0.9446\\
Entangle&0.9841&0.9431&0.9515&0.9549&0.9788

\end{tabular}
\end{ruledtabular}
\end{table}

\section{CONCLUSIONS}
We have described an experimentally feasible scheme to exponentially enhance 
the SM and OM coupling rates simultaneously through modifying a cantilever's spring constant with a driving voltage. We have introduced an electrically-driven photon-spin quantum interface for quantum networks. Even in  certain weak-coupling regime we can still design  driving  voltages to accomplish quantum state transfer and quantum entanglement distribution between two remote quantum nodes with high fidelity. Other than on NV spins  the qubits may also be encoded on  charge degrees of freedom \cite{mdla} .  The method may also find various applications in such as single-photon transistors \cite{dech}, on-demand single-photon sources \cite{blmo}, and precise measurement of optically nonactive quantum systems \cite{drug,kjhe}.

\begin{acknowledgments}
 This work was supported by the National Natural Science Foundation of China (Grant No. 11872335) and by Zhejiang Provincial Natural Science Foundation of China (Grant No. Y6110314).
\end{acknowledgments}

\end{document}